\documentclass[preprint,twocolumn]{aastex63}

\usepackage[T1]{fontenc}
 \usepackage{pslatex}
\usepackage{amsmath}
\usepackage{amsfonts}
\usepackage{color}
\usepackage{url}
\usepackage{graphicx}
\usepackage{subfigure}
\usepackage{amssymb}
\usepackage{soul}
\usepackage{booktabs}
\usepackage{multirow}
\usepackage{array}
\usepackage[utf8]{inputenc}
\inputencoding{latin1}
\inputencoding{utf8}

\usepackage{bm}
\expandafter\ifx\csname package@font\endcsname\relax\else
 \expandafter\expandafter
 \expandafter\usepackage
 \expandafter\expandafter
 \expandafter{\csname package@font\endcsname}
\fi
\expandafter\ifx\csname package@font\endcsname\relax\else
 \expandafter\expandafter
 \expandafter\usepackage
 \expandafter\expandafter
 \expandafter{\csname package@font\endcsname}
\fi
\def\bi{\begin{itemize}}
\def\ei{\end{itemize}}
\def\bq{\begin{equation}}
\def\eq{\end{equation}}
\def\bqy{\begin{eqnarray}}
\def\eqy{\end{eqnarray}}

\def\d{{\rm d}}
\def\e{{\rm e}}
\def\i{{\rm i}}
\def\pa{\partial}
\def\E{{\cal E}}

\begin{document}

\title{Saturation of the filamentation instability and dispersion measure of Fast Radio Bursts}
\shorttitle{Dispersion measure of FRBs}

\correspondingauthor{Emanuele Sobacchi}
\email{emanuele.sobacchi@mail.huji.ac.il}

\shortauthors{Sobacchi et al.}

\author{Emanuele Sobacchi}
\affiliation{Department of Astronomy and Columbia Astrophysics Laboratory, Columbia University, New York, NY 10027, USA}
\affiliation{Racah Institute for Physics, The Hebrew University, Jerusalem 91904, Israel}

\author{Yuri Lyubarsky}
\affiliation{Physics Department, Ben-Gurion University, P.O.B. 653, Beer-Sheva 84105, Israel}

\author{Andrei M. Beloborodov}
\affiliation{Physics Department and Columbia Astrophysics Laboratory, Columbia University, 538 West 120th Street New York, NY 10027, USA}
\affiliation{Max Planck Institute for Astrophysics, Karl-Schwarzschild-Str. 1, D-85741, Garching, Germany}

\author{Lorenzo Sironi}
\affiliation{Department of Astronomy and Columbia Astrophysics Laboratory, Columbia University, New York, NY 10027, USA}

\author{Masanori Iwamoto}
\affiliation{Faculty of Engineering Sciences, Kyushu University, 6-1, Kasuga-koen, Kasuga, Fukuoka, 816-8580, Japan}

\begin{abstract}
Nonlinear effects are crucial for the propagation of Fast Radio Bursts (FRBs) near the source. We study the filamentation of FRBs in the relativistic winds of magnetars, which are commonly invoked as the most natural FRB progenitors. As a result of filamentation, the particle number density and the radiation intensity develop strong gradients along the direction of the wind magnetic field. A steady state is reached when the plasma pressure balances the ponderomotive force. In such a steady state, particles are confined into periodically spaced thin sheets, and electromagnetic waves propagate between them as in a waveguide. We show that: (i) The dispersion relation resembles that in the initial homogeneous plasma, but the effective plasma frequency is determined by the separation of the sheets, not directly by the mean particle density. (ii) The contribution of relativistic magnetar winds to the dispersion measure of FRBs could be several orders of magnitude larger than previously thought. The dispersion measure of the wind depends on the properties of individual bursts (e.g. the luminosity), and therefore can change significantly among different bursts from repeating FRBs. (iii) Induced Compton scattering is suppressed because most of the radiation propagates in near vacuum regions.
\end{abstract}


\section{Introduction}

Fast Radio Bursts (FRBs) are bright flashes of millisecond duration likely produced by magnetars \citep[e.g.][]{CordesChatterjee2019, Petroff+2019, Petroff+2022}. Since near the source the electromagnetic field of the radio wave accelerates the electrons up to a significant fraction of the speed of light, nonlinear effects are crucial for the propagation of FRBs \citep[e.g.][]{Lyubarsky2021}.

In our previous work \citep[][]{Sobacchi+2022}, we studied the filamentation of FRBs in magnetar winds.\footnote{Filamentation of strong electromagnetic waves propagating in electron-ion plasmas has been extensively studied in the field of laser plasma interaction \citep[e.g.][]{Kruer2019}. Filamentation has been observed in numerical simulations of relativistic magnetised shocks, in which case a strong electromagnetic precursor is emitted upstream \citep[][]{Iwamoto+2017, Iwamoto+2022, BabulSironi2020, Sironi+2021}.} We showed that the radiation intensity and the particle density develop spatial modulations along the direction of the wind magnetic field. However, since we focused on the linear stability analysis of the filamentation instability, we could not predict what happens after the initial phase of exponential growth.

In this paper, we present a model for the saturation of the filamentation instability of FRBs in magnetar winds. Our model is inspired by classical studies of the filamentation instability in unmagnetised electron-ion plasmas \citep[][]{Kaw+1973, Max1976}. We argue that the system reaches a steady state where the ponderomotive force produced by the spatial modulations of the radiation intensity is balanced by the plasma pressure. In such a steady state, particles are confined into periodically spaced thin sheets located where the radiation intensity nearly vanishes. Electromagnetic waves propagate between the sheets as in a waveguide.

The confinement of the particles into thin sheets has important implications for FRBs: (i) The dispersion relation of electromagnetic waves resembles that in the initial homogeneous plasma, but the effective plasma frequency is determined by the separation of the sheets. (ii) The contribution of the magnetar wind to the dispersion measure of FRBs could be much larger than previously thought. (iii) The rate of induced Compton scattering is suppressed.

The paper is organised as follows. In Section \ref{sec:FRB} we review the properties of FRBs propagating in magnetar winds. In Section \ref{sec:filamentation} we study the saturation of the filamentation instability in magnetised pair plasmas. In Section \ref{sec:implications} we discuss the implications of our results for FRBs. In Section \ref{sec:conclusions} we conclude.

\section{Fast Radio Bursts in magnetar winds}
\label{sec:FRB}

\subsection{Mean particle density}

Magnetic field lines anchored to the magnetar surface open at the light cylinder, forming an electron-positron wind at radii $R\gtrsim R_{\rm LC}=cP/2\pi$ ($P$ is the magnetar rotational period). For a quiescent magnetar, one can estimate the particle outflow rate as $\dot{N}\sim 10^{39}{\rm\; s}^{-1}$, and the wind bulk Lorentz factor as $\gamma_{\rm w}\sim 10^2$ \citep[][]{Beloborodov2020}. An actively flaring magnetar is likely to produce a denser wind with a Lorentz factor as low as $\gamma_{\rm w}\sim 10$. We will keep track of how these parameters enter the final results.

The particle number density in the wind proper frame, $n_0=\dot{N}/4\pi\gamma_{\rm w}R^2c$, is
\begin{equation}
\label{eq:nwind}
n_0 = 3\times 10^{-3}\;\dot{N}_{39}\gamma_2^{-1}R_{14}^{-2}{\rm\; cm}^{-3} \;,
\end{equation}
where $\gamma_2\equiv\gamma_{\rm w}/10^2$, $\dot{N}_{39}\equiv\dot{N}/10^{39}{\rm\; s}^{-1}$, and $R_{14}\equiv R/10^{14}{\rm\; cm}$. The ratio of the plasma frequency, $\omega_{\rm P}=\sqrt{4\pi n_0e^2/m}$, and the FRB wave frequency in the wind frame, $\omega_0=\pi\nu_{\rm 0,obs}/\gamma_{\rm w}$, is
\begin{equation}
\label{eq:omegaratio}
\frac{\omega_{\rm P}}{\omega_0}=9\times 10^{-5} \dot{N}_{39}^{1/2} \gamma_2^{1/2}\nu_9^{-1}R_{14}^{-1}\;,
\end{equation}
where $\nu_9\equiv \nu_{\rm 0,obs}/1{\rm\; GHz}$ is the characteristic observed frequency in GHz units.

\subsection{Mean temperature}

Observations of weak FRBs from the Galactic magnetar SGR 1935+2154 show that radio bursts are accompanied by powerful X-ray flares \citep[][]{Chime2020,Bochenek+2020,Mereghetti+2020}. The isotropic equivalent of the observed luminosity is $L_{\rm r}\sim 10^{37}{\rm\; erg\; s}^{-1}$ for the radio, and $\mathcal{L}_{\rm x}\sim 10^{40}{\rm\; erg\; s}^{-1}$ for the X-rays. Cosmological FRBs have larger radio luminosities, $L_{\rm r}\sim 10^{42}{\rm\; erg\; s}^{-1}$. However, X-ray flares from cosmological FRBs could be hardly detected because of the enormous distance. Below we assume that the ratio of the radio to X-ray luminosities is the same as for the Galactic magnetar SGR 1935+2154, which gives $\mathcal{L}_{\rm x}\sim 10^{45}{\rm\; erg\; s}^{-1}$.

The electron temperature in the wind proper frame, $T_0$, is controlled by adiabatic expansion and by Compton heating from X-rays emitted by the magnetar.\footnote{We neglect other potential sources of heating, such as magnetic reconnection \citep[e.g.][]{LyubarskyKirk2001}, and internal shocks \citep[e.g.][]{Beloborodov2020}. If these processes were energetically important, our final estimate of the wind temperature, Eq. \eqref{eq:Twind}, would be rather a lower limit. Moreover, the plasma pressure could become dominated by non-thermal particles. Then the particle velocity distribution would not be described by a non-relativistic Maxwellian.} One may write
\begin{equation}
\label{eq:Tevo}
\frac{{\rm d} T_0}{{\rm d}R} = \left(\frac{{\rm d} T_0}{{\rm d}R}\right)_{\rm ad} + \left(\frac{{\rm d} T_0}{{\rm d}R}\right)_{\rm C}\;,
\end{equation}
where the first term on the right hand side describes adiabatic cooling, and the second term describes Compton heating.

In magnetised plasmas, the thermal velocity in the direction parallel to the local magnetic field could be different than in the perpendicular direction. The filamentation instability is independent of the thermal velocity in the perpendicular direction because particles move along the magnetic field lines as the density modulations are formed. Since the thermal velocity along the magnetic field is inversely proportional to the radius as the plasma expands \citep[][]{Chew+1956}, one may write
\begin{equation}
\label{eq:Tevoad}
\left(\frac{{\rm d} T_0}{{\rm d}R}\right)_{\rm ad} = -2\frac{T_0}{R} \;.
\end{equation}

The Compton heating rate of the electrons in the magnetar wind by the X-rays emitted near the surface is
\begin{equation}
\left(\frac{{\rm d} T_0}{{\rm d}R}\right)_{\rm C} = \frac{1}{\gamma_{\rm w}c}\dot{N}_{\rm x} \Delta T_0 \;,
\end{equation}
where
\begin{equation}
\dot{N}_{\rm x}=\frac{\sigma_{\rm T}\mathcal{L}_{\rm x}}{4\pi R^2\gamma_{\rm w}\varepsilon_{\rm x}}
\end{equation}
is the scattering rate of the X-ray photons by an electron measured in the wind proper frame ($\varepsilon_{\rm x}$ is the typical X-ray photon energy in the observer's frame), and
\begin{equation}
\label{eq:deltaT}
k_{\rm B}\Delta T_0= \frac{\varepsilon_{\rm x}}{2\gamma_{\rm w}mc^2}\left(\frac{\varepsilon_{\rm x}}{2\gamma_{\rm w}}-4k_{\rm B}T_0\right)
\end{equation}
is the energy gained by the electron in one scattering \citep[][]{RybickiLightman1979}.

Using Eqs. \eqref{eq:Tevoad}-\eqref{eq:deltaT}, Eq. \eqref{eq:Tevo} can be presented as
\begin{equation}
\label{eq:Tevo2}
\frac{{\rm d} T_0}{{\rm d}R} = -2\frac{T_0}{R} -2\left(\frac{R_{\rm cr}}{R}\right)\frac{T_0}{R} + \frac{1}{4}\left(\frac{R_{\rm cr}}{R}\right)\frac{\varepsilon_{\rm x}}{k_{\rm B}\gamma_{\rm w}R} \;,
\end{equation}
where
\begin{equation}
R_{\rm cr} = \frac{\sigma_{\rm T}\mathcal{L}_{\rm x}}{4\pi\gamma_{\rm w}^3mc^3} \;.
\end{equation}
At radii $R\ll R_{\rm cr}$, the second and the third term on the right hand side of Eq. \eqref{eq:Tevo2} are much larger than the other terms. Then Compton scattering maintains the equilibrium of the radiation and electron temperatures, and one finds $k_{\rm B}T_0=\varepsilon_{\rm x}/8\gamma_{\rm w}$. At radii $R\gg R_{\rm cr}$, the first term on the right hand side of Eq. \eqref{eq:Tevo2} is much larger than the second term. Taking into account the effect of electron recoil, one finds the solution $k_{\rm B}T_0=(\varepsilon_{\rm x}/4\gamma_{\rm w})(R_{\rm cr}/R)$. The plasma is hotter than for a pure adiabatic expansion, which would give $T_0\propto R^{-2}$. The solution of Eq. \eqref{eq:Tevo2} can be finally approximated as
\begin{equation}
k_{\rm B}T_0 =
\begin{cases}
\frac{\varepsilon_{\rm x}}{8\gamma_{\rm w}} & {\rm for}\quad R\ll R_{\rm cr} \\
\left(\frac{R_{\rm cr}}{R}\right)\frac{\varepsilon_{\rm x}}{4\gamma_{\rm w}} & {\rm for}\quad R\gg R_{\rm cr}
\end{cases}\;.
\end{equation}
At large radii $R\gg R_{\rm cr}=2\times 10^{9}\mathcal{L}_{45}\gamma_2^{-3}{\rm\; cm}$, one finds
\begin{equation}
\label{eq:Twind}
\frac{k_{\rm B}T_0}{mc^2}= 10^{-8}\mathcal{L}_{45}\varepsilon_{5}\gamma_{2}^{-4}R_{14}^{-1} \;,
\end{equation}
where $\mathcal{L}_{45}\equiv \mathcal{L}_{\rm x}/ 10^{45}{\rm\; erg\; s}^{-1}$, and $\varepsilon_5\equiv\varepsilon_{\rm x}/100{\rm\; keV}$.

\subsection{Strength parameter of the FRB wave}

The plasma density and temperature are strongly modified with respect to the mean values of Eqs. \eqref{eq:nwind} and \eqref{eq:Twind} as a result of the filamentation instability. The instability occurs because the electrons oscillate with large velocities in the electromagnetic field of the FRB wave that propagates through the wind.

The peak velocity of an electron oscillating in the electromagnetic field of the wave is $a_0c$ for $a_0\ll 1$, and becomes ultrarelativistic for $a_0\gg 1$, where the wave strength parameter is defined as $a_0=eE_0/\omega_0 mc$. The peak electric field of the wave in the wind frame, $E_0$, can be calculated from the isotropic equivalent of the observed radio luminosity, $L_{\rm r}=2c\gamma_{\rm w}^2E_0^2R^2$. The angular frequency of the wave in the wind frame is $\omega_0=\pi\nu_{\rm obs}/\gamma_{\rm w}$, where $\nu_{\rm obs}$ is the observed frequency. One finds
\begin{equation}
\label{eq:a0}
a_0=0.2\;L_{42}^{1/2}\nu_9^{-1}R_{14}^{-1}\;,
\end{equation}
where $L_{42}\equiv L_{\rm r}/10^{42}{\rm\; erg\; s}^{-1}$.

Below we consider the regime of large wave frequencies, $\omega_0\gg\omega_{\rm P}/a_0$. We restrict our study to the regime of nonrelativistic electron velocities ($a_0\ll 1$ and $k_{\rm B}T_0/mc^2\ll 1$), as appropriate for radii $R\gtrsim 10^{14}{\rm\; cm}$.

\section{Filamentation of strong waves}
\label{sec:filamentation}

\subsection{Wave equation}

We consider an electromagnetic wave propagating in the $z$ direction through a magnetised pair plasma. We focus on the case when the background magnetic field is transverse to the wave propagation (we set it to be in the $x$ direction), as appropriate in magnetar winds where the wave propagates radially and the background magnetic field is nearly azimuthal.

We look for a steady state where the large scale modulations of the radiation intensity are independent of time. Such a steady state can be achieved as the e-folding time of the filamentation instability is $\sim 100$ times shorter than the duration of the millisecond radio pulse \citep[][]{Sobacchi+2022}. Since planes of constant radiation intensity are perpendicular to the direction of the background magnetic field, the electromagnetic vector potential can be written as
\begin{equation}
\label{eq:A}
A\left({\bf x},t\right)=\frac{mc^2}{e}a\left(x\right)\int{\rm d}k_z\hat{f}\left(k_z\right)\exp\left[{\rm i}\left(k_z z-\omega t\right)\right]\;,
\end{equation}
where $\hat{f}(k_z)$ is peaked about the wave number $k_z=k_0$, and $a(x)$ describes modulations of the radiation intensity on scales $\gg k_0^{-1}$. Neglecting relativistic corrections to the electron mass, as appropriate in the limit $a_0\ll 1$, the wave equation can be written as
\begin{equation}
\label{eq:wave}
\frac{\pa^2 A}{\pa t^2}-c^2\nabla^2A+\omega_{\rm P}^2\frac{n}{n_0}A=0\;,
\end{equation}
where $\omega_{\rm P}=\sqrt{4\pi n_0e^2/m}$ is the mean plasma frequency, and $n(x)$ is the perturbed particle density. Substituting Eq. \eqref{eq:A} into Eq. \eqref{eq:wave}, for a given Fourier mode one finds
\begin{equation}
\label{eq:wave2}
\frac{{\rm d}^2 a}{{\rm d}x^2}+\left(\frac{\omega^2}{c^2}-k_z^2-\frac{\omega_{\rm P}^2}{c^2}\frac{n}{n_0}\right)a=0\;.
\end{equation}
In the next section we express $n/n_0$ as a function of $a$. Then we solve Eq. \eqref{eq:wave2} and find the dispersion relation $\omega(k_z)$.

\subsection{Particle density}

We assume that electrons and positrons have a Maxwell-Boltzmann distribution with pressure $p_0(n/n_0)^\Gamma$, where $p_0 = k_{\rm B}n_0T_0$ is the mean pressure, and $\Gamma$ is the adiabatic index. Assuming that the particles move in one dimension along the strong background magnetic field lines, one finds $\Gamma=3$.

The filamentation instability saturates when the plasma pressure balances the ponderomotive force \citep[][]{Kaw+1973, Max1976}, namely
\begin{equation}
\label{eq:equilibrium}
\frac{\Gamma}{\Gamma-1}k_{\rm B}T_0\nabla\left(\frac{n}{n_0}\right)^{\Gamma-1}+\nabla\phi=0\;,
\end{equation}
where $\phi=e^2\langle A^2\rangle/4mc^2$ is the ponderomotive potential. The time average, $\langle\ldots\rangle$, is taken on time scales $\gg \omega_0^{-1}$ \citep[][]{Chen1974}. Substituting Eq. \eqref{eq:A} into the definition of $\phi$, and choosing the arbitrary normalisation of the function $\hat{f}$ so that $\int\vert\hat{f}(k_z)\vert^2{\rm d}k_z=1$, one finds
\begin{equation}
\label{eq:phi}
\phi= \frac{1}{4}mc^2a^2\;.
\end{equation}
In cold plasmas with $\beta_{\rm s}\ll\sqrt{a_0\omega_{\rm P}/\omega_0}$, the e-folding time of the filamentation instability is shorter than the sound crossing time of the density filaments \citep[][]{Ghosh+2022, Sobacchi+2022}. In this case pressure equilibrium cannot be maintained while the instability develops, leading to supersonic bulk motions and possibly to non-adiabatic heating. Then the plasma temperature in the core of the filaments may be enhanced with respect to our estimate in Eq. \eqref{eq:betamax} below. Numerical simulations would be required to investigate this effect.

Eq. \eqref{eq:equilibrium} should be satisfied only in the regions where $n\neq 0$, since the ponderomotive force does not have any particles to push where $n=0$. Substituting Eq. \eqref{eq:phi} into Eq. \eqref{eq:equilibrium}, one finds
\begin{equation}
\label{eq:n}
\frac{n}{n_0}= 
\begin{cases}
\left(\frac{\beta_{\rm max}^2-a^2}{\beta_{\rm s}^2}\right)^{\frac{1}{\Gamma-1}} & {\rm for}\quad a<\beta_{\rm max} \\
0 & {\rm for}\quad a>\beta_{\rm max}
\end{cases}\;,
\end{equation}
where
\begin{equation}
\label{eq:betadef}
\beta_{\rm s}=\sqrt{\frac{4\Gamma}{\Gamma-1}\frac{k_{\rm B}T_{\rm 0}}{mc^2}}
\end{equation}
is the mean thermal velocity along the background magnetic field, and
\begin{equation}
\label{eq:betamax}
\beta_{\rm max}=\left(\frac{n_{\rm max}}{n_0}\right)^{\frac{\Gamma-1}{2}}\beta_{\rm s}
\end{equation}
is the peak thermal velocity that is achieved for $a=0$ (in the core of the density sheets).

\subsection{Solution of the wave equation}

Eqs. \eqref{eq:wave2} and \eqref{eq:n} provide a second order differential equation for $a(x)$, which is a periodic function\footnote{As discussed in Appendix \ref{sec:appA}, there are also cavity solutions where the radiation intensity is confined in a single slab. Cavity solutions could describe the self-focusing of a laser beam \citep[][]{Kaw+1973, Max1976}. However, the FRB wave front is broken into a large number of filaments (rather than being focused as a whole) because the transverse size of the beam is much larger than the spatial scale of the radiation intensity modulations. In this case, periodic solutions are more relevant.} of $x$. The solution is subject to two constraints (Eqs. \eqref{eq:ncons} and \eqref{eq:econs} below).

The total number of particles should be the same as in the unperturbed solution (where $n(x)=n_0$ and $a(x)=a_0$ are equal to the mean values). This condition requires
\begin{equation}
\label{eq:ncons}
\frac{1}{x_0}\int_0^{x_0}n\left(x\right){\rm d}x=n_0\;,
\end{equation}
where $x_0$ is the period of $n(x)$. 

The total energy density should be the same as in the unperturbed solution. The ratio of the electromagnetic energy density, $\omega_0^2m^2c^2a_0^2/e^2$, and the particle thermal energy density, $n_0m\beta_{\rm s}^2c^2$, is $a_0^2\omega_0^2/\beta_{\rm s}^2\omega_{\rm P}^2$. For large wave frequencies $\omega_0\gg\omega_{\rm P}/a_0$, the energy density is dominated by the electromagnetic fields. Then the conservation of the total energy requires
\begin{equation}
\label{eq:econs}
\frac{1}{x_0}\int_0^{x_0}a^2\left(x\right){\rm d}x=a_0^2\;.
\end{equation}
Below we solve Eqs. \eqref{eq:wave2} and \eqref{eq:n} for a cold plasma with $\beta_{\rm s}\ll a_0$, and a for hot plasma with $\beta_{\rm s}\gg a_0$.

\subsubsection{Cold plasma ($\beta_{\rm s}\ll a_0$)}
\label{sec:cold}

\begin{figure}{\vspace{3mm}} 
\centering
\includegraphics[width=0.46\textwidth]{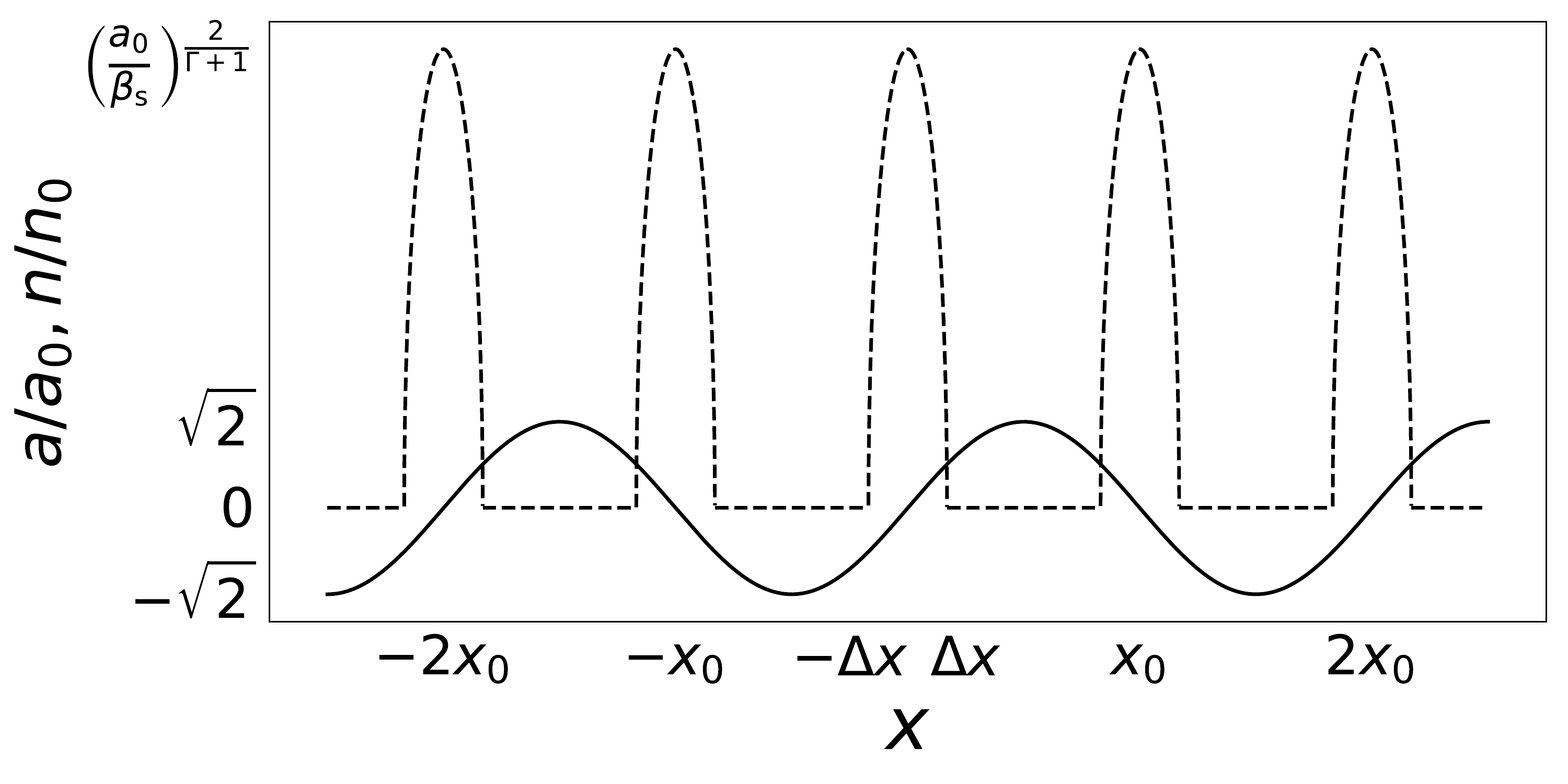}
\vspace{-3mm}
\caption{Sketch of the electromagnetic vector potential $a$ (solid) and particle number density $n$ (dashed) in the plane perpendicular to the direction of propagation of the wave. The magnetic field of the magnetar wind is directed along $x$. We consider a cold plasma with $\beta_{\rm s}\ll a_0$, and large wave frequencies $\omega_0\gg\omega_{\rm P}/a_0$.
}
\label{fig:modulations}
\end{figure}

\begin{table*}
\begin{center}
\begin{tabular}{cccccc}
\hline \\ [-2.4ex] 
range of $\beta_{\rm s}$ & $n_{\rm max}/n_0$ & $\beta_{\rm max}$ & $x_0$ & $\Delta x/x_0$ & $\omega_{\rm P,eff}$ \\ [0.5ex] 
\hline \\ [-2.2ex]
$\sqrt{a_0\omega_{\rm P}/\omega_0}\ll\beta_{\rm s}\ll a_0$ & \multirow{2}{*}{$\left(a_0/\beta_{\rm s}\right)^{2/\left(\Gamma+1\right)}$} & \multirow{2}{*}{$\left(\beta_{\rm s}/a_0\right)^{2/\left(\Gamma+1\right)}a_0$} & $x_0\sim \beta_{\rm s}c/a_0\omega_{\rm P}$ & \multirow{2}{*}{$\left(\beta_{\rm s}/a_0\right)^{2/\left(\Gamma+1\right)}$} & \multirow{2}{*}{$c/x_0$} \\ [0.5ex]
$\beta_{\rm s}\ll\sqrt{a_0\omega_{\rm P}/\omega_0}$ & & & $\beta_{\rm s}c/a_0\omega_{\rm P}\lesssim x_0\lesssim c/\sqrt{a_0\omega_0\omega_{\rm P}}$ & & \\ [0.5ex]
\hline
\end{tabular}
\caption{\label{table} Peak particle number density ($n_{\rm max}$), peak thermal velocity ($\beta_{\rm max}$), separation of the density sheets ($x_0$), thickness of the sheets ($\Delta x$), and effective plasma frequency ($\omega_{\rm P,eff}$). We consider a cold plasma with $\beta_{\rm s}\ll a_0$, and large wave frequencies $\omega_0\gg\omega_{\rm P}/a_0$. We neglect numerical factors of order unity.}
\end{center}
\end{table*}

When $\beta_{\rm s}\ll a_0$, the ponderomotive potential, $\phi\sim mc^2a_0^2$, is much larger than the thermal energy per particle, $k_{\rm B}T_0\sim mc^2\beta_{\rm s}^2$. Since in Eq. \eqref{eq:equilibrium} the ponderomotive force is balanced by the plasma pressure, the particle number density should vary on very short spatial scales. This leads to the confinement of the particles into thin sheets located where the ponderomotive force nearly vanishes.

The solution of Eq. \eqref{eq:wave2} can be determined as follows. Since particles are confined into thin sheets, and the particle density vanishes practically everywhere, the last term of Eq. \eqref{eq:wave2} can be neglected (a formal justification of our assumption is presented in Appendix \ref{sec:appB}). Then Eq. \eqref{eq:wave2} becomes
\begin{equation}
\label{eq:wavecold}
\frac{{\rm d}^2 a}{{\rm d}x^2}+\left(\frac{\omega^2}{c^2}-k_z^2\right)a=0\;.
\end{equation}
The solution of Eq. \eqref{eq:wavecold} can be presented as
\begin{equation}
\label{eq:acold}
a=\sqrt{2}a_0\sin\left(\pi\frac{x}{x_0}\right)\;,
\end{equation}
where
\begin{equation}
\label{eq:DRcold}
\omega^2=c^2k_z^2+\left(\frac{\pi c}{x_0}\right)^2\;.
\end{equation}
The factor $\sqrt{2}a_0$ comes from the conservation of the total energy, Eq. \eqref{eq:econs}. The dispersion relation resembles that in the homogeneous plasma, with an effective plasma frequency
\begin{equation}
\label{eq:omegaeff}
\omega_{\rm P,eff}=\frac{\pi c}{x_0}\;.
\end{equation}
This result has a simple physical interpretation \citep[][]{Max1976}. Eq. \eqref{eq:DRcold} is analogous to the dispersion relation of electromagnetic waves propagating in a waveguide. Since the electromagnetic fields vanish for $x=0$ and $x=x_0$, the cutoff frequency is $\pi c/x_0$.

Particles are confined into thin sheets located where $a\sim 0$. The half thickness of the sheets, $\Delta x$, can be determined from Eq. \eqref{eq:n}, which gives $a(\Delta x)= \beta_{\rm max}$. Approximating Eq. \eqref{eq:acold} for small arguments of the sine function, $a(\Delta x)\sim \sqrt{2}\pi a_0\Delta x/x_0$, one finds
\begin{equation}
\label{eq:deltax1}
\frac{\Delta x}{x_0}\sim \frac{\beta_{\rm max}}{a_0} \;.
\end{equation}
The conservation of the total number of particles, Eq. \eqref{eq:ncons}, gives
\begin{equation}
\label{eq:nmax2}
n_{\rm max}\Delta x \sim n_0x_0 \;.
\end{equation}
From Eqs. \eqref{eq:betamax}, \eqref{eq:deltax1}, and \eqref{eq:nmax2}, one finds
\begin{equation}
\label{eq:acr}
\frac{\Delta x}{x_0} \sim \frac{n_0}{n_{\rm max}} \sim \frac{\beta_{\rm max}}{a_0} \sim \left(\frac{\beta_{\rm s}}{a_0}\right)^{\frac{2}{\Gamma+1}} \;.
\end{equation}
The electromagnetic vector potential, $a(x)$, and the particle number density, $n(x)$, are sketched in Figure \ref{fig:modulations}.

The spatial scale of the radiation intensity modulations, $x_0$, can be estimated as follows. Linear stability analysis shows that the unperturbed solution is unstable \citep[][]{Ghosh+2022, Sobacchi+2022}. When $\sqrt{a_0\omega_{\rm P}/\omega_0}\ll \beta_{\rm s}\ll a_0$, the growth rate of filamentation is sharply peaked about $k_x=a_0\omega_{\rm P}/2\beta_{\rm s}c$. On the other hand, when $\beta_{\rm s}\ll\sqrt{a_0\omega_{\rm P}/\omega_0}$, all the wave numbers $\sqrt{a_0\omega_0\omega_{\rm P}}/c \lesssim k_x\lesssim a_0\omega_{\rm P}/\beta_{\rm s}c$ grow practically at the same rate. Extrapolating the results of the linear stability analysis to the nonlinear stage, and assuming $x_0\sim \pi/k_x$, one finds\footnote{Since a wide range of wave numbers grow practically at the same rate when $\beta_{\rm s}\ll\sqrt{a_0\omega_{\rm P}/\omega_0}$, our argument gives only lower and upper limits of $x_0$. To determine the actual $x_0$, one should study how the instability saturates for various wave numbers, which is out of the scope of the paper. We speculate that $x_0$ is determined by the power spectrum of the initial intensity fluctuations from which the instability develops. If the initial power spectrum peaks at large wave numbers, one may have $x_0 \sim \beta_{\rm s}c/a_0\omega_{\rm P}$. If the initial power spectrum peaks at small wave numbers, one may have $x_0 \sim c/\sqrt{a_0\omega_0\omega_{\rm P}}$.}
\begin{align}
\label{eq:x01}
x_0 & \sim \frac{\beta_{\rm s}c}{a_0\omega_{\rm P}} & {\rm for} & \quad \sqrt{a_0\omega_{\rm P}/\omega_0}\ll\beta_{\rm s}\ll a_0 \\
\label{eq:x02}
\frac{\beta_{\rm s}c}{a_0\omega_{\rm P}}\lesssim x_0 & \lesssim \frac{c}{\sqrt{a_0\omega_0\omega_{\rm P}}} & {\rm for} & \quad \sqrt{a_0\omega_{\rm P}/\omega_0}\gg\beta_{\rm s} \;.
\end{align}
Our results are summarised in Table \ref{table}. For large wave frequencies $\omega_0\gg\omega_{\rm P}/a_0$, the effective plasma frequency $\omega_{\rm P,eff}=\pi c/x_0$ is much larger than $\omega_{\rm P}$.

The above estimates predict the formation of dense plasma sheets separated by near vacuum regions. Similar structures were observed in numerical simulations of a strong electromagnetic wave propagating in a cold magnetised pair plasma \citep[][]{Sironi+2021}. The sheets were found to persist on the timescale of the simulations ($\sim 10^4/\omega_{\rm P}$).

\subsubsection{Hot plasma ($\beta_{\rm s}\gg a_0$)}

When $\beta_{\rm s}\gg a_0$, the ponderomotive potential is much smaller than the thermal energy per particle. Then  the ponderomotive force is balanced by small fluctuations of the particle density, $|n-n_0|\ll n_0$, even for large modulations of the radiation intensity.

The dispersion relation can be determined as follows. The ratio of the first and the last term of Eq. \eqref{eq:wave2} is $(c^2/\omega_{\rm P}^2x_0^2)(n_0/n)\lesssim a_0^2/\beta_{\rm s}^2\ll 1$. Neglecting the first term, one finds
\begin{equation}
\label{eq:DRhot}
\omega^2=c^2k_z^2+\omega_{\rm P}^2\;.
\end{equation}
The dispersion relation is not modified by filamentation.

\section{Implications for Fast Radio Bursts}
\label{sec:implications}

\subsection{Effective dispersion measure}

Below we calculate the effective dispersion measure of FRBs propagating in magnetar winds. As discussed in Section \ref{sec:filamentation}, the effective plasma frequency $\omega_{\rm P,eff}$ for a cold plasma is much larger than $\omega_{\rm P}$ as an effect of filamentation. Then the dispersion measure is much larger than the value obtained for a uniform plasma density.

From Eq. \eqref{eq:DRcold} and \eqref{eq:omegaeff}, one can calculate the group velocity of electromagnetic waves in the wind proper frame,
\begin{equation}
\label{eq:vg}
v_{\rm g}=c\sqrt{1-\frac{\omega_{\rm P,eff}^2}{\omega^2}} \;.
\end{equation}
The group velocity in the observer's frame is
\begin{equation}
\label{eq:vgobs}
v_{\rm g,obs}=\frac{v_{\rm g}+v_{\rm w}}{1+v_{\rm g}v_{\rm w}/c^2}\;,
\end{equation}
where $v_{\rm w}$ is the velocity of the wind. The correction to the light travel time can be presented as
\begin{equation}
\label{eq:tw}
t_{\rm w}=\int \frac{{\rm d}R}{c}\left(\frac{c}{v_{\rm g,obs}}-1\right)\simeq \frac{1}{2}\int\frac{{\rm d}R}{c}\frac{\omega_{\rm P,eff}^2}{\omega_{\rm obs}^2} \;,
\end{equation}
where $\omega_{\rm obs}=2\gamma_{\rm w}\omega$ is the angular frequency in the observer's frame.

Unless magnetic reconnection or internal shocks heat the plasma well above the fiducial temperature of Eq. \eqref{eq:Twind}, one has $\beta_{\rm s}\lesssim\sqrt{a_0\omega_{\rm P}/\omega_0}$, where $\omega_0$ denotes the power-weighted angular frequency of the burst in the wind frame. The effective plasma frequency in the wind proper frame can be estimated from Eqs. \eqref{eq:omegaeff} and \eqref{eq:x02}, which give $\sqrt{a_0 \omega_0 \omega_{\rm P}} \lesssim\omega_{\rm P,eff}\lesssim (a_0/\beta_{\rm s})\omega_{\rm P}$. Then the contribution of the wind to the dispersion measure is $\int(a_0\omega_0/\omega_{\rm P}) n_0{\rm d}R \lesssim {\rm DM}_{\rm w}\lesssim\int(a_0/\beta_{\rm s})^2n_0{\rm d}R$. Using Eqs. \eqref{eq:nwind}, \eqref{eq:omegaratio}, \eqref{eq:Twind}, and \eqref{eq:a0}, one finds
\begin{equation}
\label{eq:DM}
{\rm DM}_{\rm w,low}\lesssim{\rm DM}_{\rm w}\lesssim{\rm DM}_{\rm w,high}\;,
\end{equation}
where
\begin{align}
{\rm DM}_{\rm w,low} & =  2\times 10^{-4}\dot{N}_{39}^{1/2}L_{42}^{1/2}\gamma_2^{-3/2}R_{14}^{-1} {\rm\; pc}{\rm\; cm}^{-3} \\
\label{eq:DMhigh}
{\rm DM}_{\rm w,high} & = 4\times 10^{-2}\dot{N}_{39}L_{42}\mathcal{L}_{45}^{-1}\varepsilon_{5}^{-1}\gamma_{2}^3\nu_9^{-2}R_{14}^{-2} {\rm\; pc}{\rm\; cm}^{-3}\;.
\end{align}
For the sake of comparison, the nominal dispersion measure of the wind (i.e. neglecting filamentation) is $\int n_0{\rm d}R= 8\times 10^{-8}\dot{N}_{39}\gamma_2^{-1}R_{14}^{-1} {\rm\; pc}{\rm\; cm}^{-3}$. Note that $\nu_9=\nu_{\rm 0,obs}/1{\rm\;GHz}$ denotes the power-weighted frequency of the burst in the observer's frame.

Eqs. \eqref{eq:DM}-\eqref{eq:DMhigh} show that the contribution of the magnetar wind to the FRB dispersion measure depends on the properties of the radio burst and of the X-ray flare.\footnote{\citet[][]{LuPhinney2020} suggested that the dispersion measure of luminous bursts may decrease due to relativistic corrections to the effective electron mass. In the magnetar wind, this effect is negligible because the inferred fluctuations of the dispersion measure are of order $\int a_0^2n_0{\rm d}R=10^{-9}\dot{N}_{39}L_{42}\gamma_2^{-1}\nu_9^{-2}R_{14}^{-3}{\rm\; pc\; cm}^{-3}$.} Then the dispersion measure can fluctuate significantly among different bursts from repeating FRBs. The amplitude of the fluctuations is of order ${\rm DM}_{\rm w}$.

Interestingly, the dispersion measure of the bursts from the repeating FRB 20190520B fluctuates by $\sim 40{\rm\; pc\; cm}^{-3}$ on time scales $\lesssim 10{\rm\; s}$. These fluctuations went unnoticed in the original paper of \citet[][]{Dai+2022}, and were subsequently pointed out by \citet[][]{Katz2022}. Fluctuations of the dispersion measure by $1-10 {\rm\; pc\; cm}^{-3}$ on time scales $\lesssim 100{\rm\; s}$ were also reported in FRB 20121102A \citep[][]{Li+2021} and FRB 20201124A \citep[][]{Xu+2022}, although their importance was not fully appreciated. The observed fluctuations could be produced by the magnetar wind at radii $R\sim 10^{14}{\rm\; cm}$. Eq. \eqref{eq:DMhigh} shows that ${\rm DM}_{\rm w,high}$ is proportional to the particle outflow rate times the wind Lorentz factor cube, $\dot{N}\gamma_{\rm w}^3$, and to the ratio of the radio to X-ray luminosities, $L_{\rm r}/\mathcal{L}_{\rm x}$. If the wind is accelerated until the fast magnetosonic point, one finds $\dot{N}\gamma_{\rm w}^3\sim 30\;L_{\rm w}/mc^2$, where $L_{\rm w}$ is the luminosity of the wind \citep[][]{Beloborodov2020}. The luminosity could be significantly larger than our fiducial value, $L_{\rm w}=3\times 10^{37}{\rm\; erg\; s}^{-1}$, thus increasing ${\rm DM}_{\rm w,high}$. The ratio of the radio to X-ray luminosities could also exceed our fiducial value, $L_{\rm r}/\mathcal{L}_{\rm x}=10^{-3}$. FRBs from the Galactic magnetar SGR 1935+2154 were preceding the X-ray flare by $\sim 5{\rm\; ms}$ \citep[][]{Mereghetti+2020}. Then the X-ray luminosity that heats the wind during the passage of the FRB could be reduced with respect to the peak value. Taking $L_{\rm w}=3\times 10^{39}{\rm\; erg\;s}^{-1}$ and $L_{\rm r}/\mathcal{L}_{\rm x}=10^{-2}$, Eq. \eqref{eq:DMhigh} gives ${\rm DM}_{\rm w,high}\sim 40 {\rm\; pc\; cm}^{-3}$, consistent with the observed fluctuations of the dispersion measure of FRB 20190520B.

Another possibility is that the plasma at small radii $R\lesssim 10^{13}{\rm\; cm}$ significantly contributes to the wind dispersion measure. However, such a contribution cannot be quantified by our estimates, which are valid only in the limit of small wave strength parameter $a_0\ll 1$ relevant for large radii (see Eq. \ref{eq:a0}).

\subsection{Suppression of induced Compton scattering}

Induced Compton scattering of FRBs has been discussed by several authors \citep[for a review, see][]{Lyubarsky2021}. Assuming the particle density and the radiation intensity to be spatially uniform, the induced scattering rate can be estimated as $\Gamma_{\rm IC}\sim a_0^2\omega_{\rm P}^2/\omega_0$. However, since induced scattering grows at a slower rate with respect to filamentation when the radio burst has a broad spectrum \citep[][]{Ghosh+2022},\footnote{This is not necessarily true in electron-ion plasmas, because the growth rate of the filamentation instability is lower due to the large inertia of the ions \citep[][]{Drake+1974, Sobacchi+2021}. Moreover, even in a cold plasma with $\beta_{\rm s}\ll a_0$ the formation of particle sheets with a large density contrast may be hindered due to the large inertia of the ions \citep[e.g.][]{SchmittAfeyan1998}.} the particle density and the radiation intensity are by no means uniform. The structures illustrated in Figure \ref{fig:modulations} are formed on a few e-folding times of the filamentation instability, well before induced scattering can operate.

The rate of induced Compton scattering is much smaller than the standard rate calculated for a uniform particle and radiation density. Since $\Gamma_{\rm IC}\propto a^2n$, the scattering rate is not spatially uniform. Outside the sheets where the particles are confined, the scattering rate vanishes because $n=0$. Inside the sheets, the particle density is $n\sim (a_0/\beta_{\rm s})^{2/(\Gamma+1)}n_0$, and the strength parameter of the wave is $a\sim (\Delta x/x_0)a_0\sim (\beta_{\rm s}/a_0)^{2/(\Gamma+1)} a_0$ (see Eq. \ref{eq:acr}). Then the scattering rate is suppressed by a factor $(a/a_0)^2 (n/n_0) \sim (\beta_{\rm s}/a_0)^{2/(\Gamma+1)}\ll 1$.

\section{Conclusions}
\label{sec:conclusions}

We studied the saturation of the filamentation instability of Fast Radio Bursts propagating in magnetised pair plasmas. Due to the instability, the particle number density and the radiation intensity are spatially modulated along the direction of the background magnetic field (assumed to be perpendicular to the direction of propagation of the wave). We find that particles are confined into periodically spaced thin sheets located where the radiation intensity nearly vanishes. Electromagnetic waves propagate between the sheets as in a waveguide. This effect has important implications:
\begin{enumerate}
\item The dispersion relation resembles that in the initial homogeneous plasma with an effective plasma frequency $\omega_{\rm P,eff}= \pi c/x_0$, where $x_0$ is the separation of the density sheets.
\item The contribution of the magnetar wind to the dispersion measure of FRBs (${\rm DM}_{\rm w}$) could be much larger than previously thought, possibly reaching a few tens of ${\rm pc\; cm}^{-3}$. Since ${\rm DM}_{\rm w}$ depends on the properties of the radio burst and of the accompanying X-ray flare, the dispersion measure can change significantly among different bursts from repeating FRBs. This effect may explain the fluctuations of the dispersion measure of the repeating FRB 20190520B.
\item The rate of induced Compton scattering is much smaller than the standard rate calculated for a uniform particle density. Outside the sheets where the particles are confined, the scattering rate vanishes because the radiation  propagates in vacuum. Inside the sheets, the scattering rate is suppressed because the radiation intensity nearly vanishes.
\end{enumerate}
The calculation of ${\rm DM}_{\rm w}$ is affected by some uncertainties: (i) The properties of the magnetar wind (bulk Lorentz factor, particle outflow rate, thermal velocity dispersion) may significantly change with time as a result of the flaring activity of the magnetar. (ii) The separation of the density sheets, $x_0$, is uncertain. Our estimate of $x_0$ relies on the extrapolation of the results of the linear stability analysis of the filamentation instability. Since the linear analysis predicts that a wide range of wave numbers grows practically at the same rate, we could only determine lower and upper limits of $x_0$. (iii) Our results are valid in the limit of nonrelativistic electron velocities (wave strength parameter $a_0\ll 1$, and thermal velocity $\beta_{\rm s}\ll 1$). The contribution of the circum-source plasma (radii $R\lesssim 10^{13}{\rm\; cm}$, where $a_0\gtrsim 1$) to ${\rm DM}_{\rm w}$ should be quantified elsewhere.

\section*{Acknowledgments}

We thank the anonymous referee for constructive comments and suggestions that improved the paper. YL acknowledges support from the Israeli Science Foundation grant 2067/19. AMB acknowledges support from the Simons Foundation grant \#446228, NSF AST-2009453, and NASA 21-ATP21-0056. LS acknowledges support from the Cottrell Scholars Award and NASA 80NSSC18K1104. MI acknowledges support from JSPS KAKENHI grant  No. 20J00280 and 20KK0064.

\appendix
%\twocolumngrid

\section{Cavity solutions}
\label{sec:appA}

Eq. \eqref{eq:wave2} can be presented as
\begin{equation}
\label{eq:motion}
\frac{1}{2}\left(\frac{{\rm d}a}{{\rm d}x}\right)^2 +V\left(a\right) = C \;,
\end{equation}
where
\begin{equation}
\label{eq:V}
V\left(a\right) = \frac{1}{2}\left(\frac{\omega^2}{c^2}-k_z^2\right)a^2 + \frac{1}{2}\frac{\Gamma-1}{\Gamma} \beta_{\rm s}^2 \frac{\omega_{\rm P}^2}{c^2}\left(\frac{n}{n_0}\right)^\Gamma\;.
\end{equation}
The particle number density $n(a)$ is given by Eq. \eqref{eq:n}, and $C$ is an integration constant. Eq. \eqref{eq:motion} is equivalent to the equation of motion of a particle in the potential $V(a)$.

In cavity solutions, the radiation is confined in a single slab. Such solutions are obtained when $a={\rm d}a/{\rm d}x=0$ for $|x|\to\infty$ \citep[][]{Kaw+1973, Max1976}. The conservation of the total number of particles requires $\beta_{\rm max}=\beta_{\rm s}$, so that $n=n_0$ for $|x|\to\infty$. Substituting $a={\rm d}a/{\rm d}x=0$ into Eq. \eqref{eq:motion}, one finds the integration constant
\begin{equation}
\label{eq:C}
C = \frac{1}{2}\frac{\Gamma-1}{\Gamma} \beta_{\rm s}^2 \frac{\omega_{\rm P}^2}{c^2} \;.
\end{equation}

The dispersion relation is determined substituting $a=a_0$ into Eq. \eqref{eq:motion}, where $a_0$ is the maximal value of $a$. Since ${\rm d}a/{\rm d}x=0$ for $a=a_0$, one finds
\begin{equation}
\label{eq:DRapp}
V\left(a_0\right)=C\;,
\end{equation}
where $V\left(a_0\right)$ and $C$ are given by Eqs. \eqref{eq:V} and \eqref{eq:C}. If $a_0\gg\beta_{\rm s}$, for $a=a_0$ one finds $n=0$. If $a_0\ll\beta_{\rm s}$, for $a=a_0$ one finds $(n/n_0)^{\Gamma}\simeq 1-[\Gamma/(\Gamma-1)]a_0^2/\beta_{\rm s}^2$. Then Eq. \eqref{eq:DRapp} gives
\begin{equation}
\omega^2=c^2k_z^2+\omega_{\rm P,eff}^2 \;,
\end{equation}
where
\begin{equation}
\omega_{\rm P,eff}^2 = 
\begin{cases}
\frac{\Gamma-1}{\Gamma}\frac{\beta_{\rm s}^2}{a_0^2}\omega_{\rm P}^2 & {\rm for}\quad a_0\gg\beta_{\rm s} \\
\omega_{\rm P}^2 & {\rm for}\quad a_0\ll\beta_{\rm s}
\end{cases}\;.
\end{equation}
When $a_0\gg\beta_{\rm s}$, the effective plasma frequency is suppressed ($\omega_{\rm P,eff}\ll\omega_{\rm P}$) because in cavity solutions the spatial scale of the region of large radiation intensity is larger than the nominal plasma skin depth $c/\omega_{\rm P}$.

\section{Derivation of Eq. (23)}
\label{sec:appB}

In order to obtain Eq. \eqref{eq:wavecold}, it is sufficient to show that the second term of the potential $V(a)$ (defined in Eq. \ref{eq:V}) can be neglected. If $V(a_0)\gg V(0)$, the second term of $V(a)$ can be neglected because it gives a small correction to the particle velocity when $a\sim 0$.

When $a\sim a_0$, one has $n=0$. Since $\omega^2/c^2-k_z^2= \pi^2/x_0^2$ (see Eq. \ref{eq:DRcold}), the value of the potential is $V(a_0)\sim a_0^2/x_0^2$. When $a\sim 0$, one has $n\sim (a_0/\beta_{\rm s})^{2/(\Gamma+1)}n_0$ (see Eq. \ref{eq:acr}). Then the value of the potential is $V(0)\sim \beta_{\rm s}^2(\omega_{\rm P}^2/c^2)(a_0/\beta_{\rm s})^{2\Gamma/(\Gamma+1)}$. The condition $V(a_0)\gg V(0)$ is satisfied for
\begin{equation}
\label{eq:x0lim}
x_0 \ll \left(\frac{a_0}{\beta_{\rm s}}\right)^{\frac{1}{\Gamma+1}}\frac{c}{\omega_{\rm P}}\;.
\end{equation}
When $\omega_0\gg\omega_{\rm P}/a_0$ and $\beta_{\rm s}\ll a_0$, from Eqs. \eqref{eq:x01} and \eqref{eq:x02} one finds $x_0\ll c/\omega_{\rm P}$. Then Eq. \eqref{eq:x0lim} is satisfied.

\bibliographystyle{aasjournal}
\bibliography{2d}

\end{document}